\documentclass[fleqn,usenatbib]{mnras}

\usepackage{newtxtext,newtxmath}

\usepackage[T1]{fontenc}

\DeclareRobustCommand{\VAN}[3]{#2}
\let\VANthebibliography\thebibliography
\def\thebibliography{\DeclareRobustCommand{\VAN}[3]{##3}\VANthebibliography}

\newcommand{\hi}{\textrm{H\textsc{i}}}
\newcommand{\secref}[1]{\hyperref[#1]{Section~\ref*{#1}}}
\newcommand{\appref}[1]{\hyperref[#1]{Appendix~\ref*{#1}}}
\usepackage{xcolor}
\newcommand{\edit}[1]{\textcolor{black}{#1}}
\newcommand{\editt}[1]{\textcolor{black}{#1}}
\newcommand{\edittt}[1]{\textcolor{black}{#1}}

\usepackage{graphicx}	
\usepackage{amsmath}	

\title[MeerKAT IM ${\times}$ WiggleZ power spectrum detection]{\hi\ intensity mapping with MeerKAT: power spectrum detection in cross-correlation with WiggleZ galaxies}

\author[Cunnington, Li et al.]{Steven Cunnington$^{1,2}$\thanks{steven.cunnington@manchester.ac.uk},
Yichao Li$^{3,4}$\thanks{liyichao@mail.neu.edu.cn},
Mario G. Santos$^{4,5}$,
Jingying Wang$^{4,6}$, 
Isabella P. Carucci$^{7,8}$,
\newauthor
Melis O. Irfan$^{4,9}$,
Alkistis Pourtsidou$^{2,10,4}$,
Marta Spinelli$^{11,4}$,
Laura Wolz$^1$,
Paula S. Soares$^{9}$,
\newauthor
Chris Blake$^{12}$,
Philip Bull$^{9,4}$,
Brandon Engelbrecht$^{4}$,
Jos\'e Fonseca$^{13}$,
Keith Grainge$^1$,
Yin-Zhe Ma$^{14,15}$
\\
$^{1}$Jodrell Bank Centre for Astrophysics, Department of Physics \& Astronomy, The University of Manchester, Manchester M13 9PL, UK\\
$^{2}$Institute for Astronomy, The University of Edinburgh, Royal Observatory, Edinburgh EH9 3HJ, UK\\
$^3$Department of Physics, College of Sciences, Northeastern University, Wenhua Road, Shenyang, 11089, China\\
$^4$Department of Physics and Astronomy, University of the Western
Cape, Robert Sobukwe Road, Cape Town, 7535, South Africa\\
$^5$South African Radio Astronomy Observatory (SARAO), 2 Fir Street, Cape Town, 7925, South Africa\\
$^{6}$Shanghai Astronomical Observatory, Chinese Academy of Sciences, 80 Nandan Road, Shanghai, 200030, China\\
$^7$Dipartimento di Fisica, Universit\`a degli Studi di Torino, via P. Giuria 1, 10125, Torino, Italy\\
$^8$INFN – Istituto Nazionale di Fisica Nucleare, Sezione di Torino, via P. Giuria 1, 10125, Torino, Italy\\
$^{9}$Department of Physics \& Astronomy, Queen Mary University of London, London, E1 4NS, UK\\
$^{10}$Higgs Centre for Theoretical Physics, School of Physics and Astronomy, The University of Edinburgh, Edinburgh EH9 3FD, UK\\
$^{11}$Institute of Particle Physics \& Astrophysics, Department of Physics, ETH Zurich, Switzerland\\
$^{12}$Centre for Astrophysics \& Supercomputing, Swinburne University of Technology, P.O. Box 218, Hawthorn, VIC 3122, Australia\\
$^{13}$Instituto de Astrofísica e Ciências do Espaço, Universidade do Porto
CAUP, Rua das Estrelas, PT4150-762 Porto, Portugal\\
$^{14}$NAOC-UKZN Computational Astrophysics Center (NUCAC), University of Kwazulu-Natal, Durban, 4000, South Africa\\
$^{15}$School of Chemistry and Physics, University of KwaZulu-Natal, Westville Campus, Private Bag X54001, Durban, South Africa
}

\date{Accepted XXX. Received YYY; in original form ZZZ}

\pubyear{2015}

\begin{document}
\label{firstpage}
\pagerange{\pageref{firstpage}--\pageref{lastpage}}
\maketitle

\begin{abstract}
We present a detection of correlated clustering between MeerKAT radio intensity maps and galaxies from the WiggleZ Dark Energy Survey. We find a $7.7\sigma$ detection of the cross-correlation power spectrum, the amplitude of which is proportional to the product of the \hi\ density fraction ($\Omega_\hi$), \hi\ bias ($b_\hi$) and the cross-correlation coefficient ($r$). We therefore obtain the constraint $\Omega_\hi b_\hi r\,{=}\,[0.86\,{\pm}\,0.10\,({\rm stat})\,{\pm}\,0.12\,({\rm sys})]\,{\times}\,10^{-3}$, at an effective scale of $k_{\rm eff}\,{\sim}\,0.13\,h\,\text{Mpc}^{-1}$. The intensity maps were obtained from a pilot survey with the MeerKAT telescope, a 64-dish pathfinder array to the SKA Observatory (SKAO). The data were collected from 10.5 hours of observations using MeerKAT's L-band receivers over six nights covering the 11hr field of WiggleZ, in the frequency range 1015--973\,MHz (0.400$\,{<}\,z\,{<}\,$0.459 in redshift). This detection is the first practical demonstration of the multi-dish auto-correlation intensity mapping technique for cosmology. This marks an important milestone in the roadmap for the cosmology science case with the full SKAO.
\end{abstract}




\section{Introduction}

Probing the large scale structure of the Universe is a crucial step towards precision cosmology as we try to constrain the nature of dark energy, non-Gaussian fluctuations in the Universe's primordial density field, and test general relativity. Typically, this is done using galaxy surveys with spectroscopic or photometric redshifts in the optical or near-infrared. At radio wavelengths, we use the redshifted neutral hydrogen (\hi) hyperfine transition line, with a rest-frame wavelength of 21cm, to measure redshift. Given the ubiquitous nature of \hi\ in the Universe, we can use it to trace the distribution of dark matter at low and high redshifts. 

The faintness of the \hi\ emission line makes it challenging to resolve individual galaxies at higher redshifts over large volumes. However, for cosmology we are interested in the bulk fluctuations on large (Mpc) scales, so we can use the \hi\ intensity mapping technique. This technique relaxes the requirement of galaxy detection by integrating all 21cm emission within relatively large spatial voxels \citep{Bharadwaj:2000av,Battye:2004re,Wyithe:2007rq,Chang:2007xk}. This delivers high survey speeds over large volumes, providing a novel solution to the current challenges of observational cosmology.

One of the main challenges in detecting the \hi\ intensity mapping signal is the presence of foregrounds that are orders of magnitude brighter. Removing these requires precise instrumental calibration. Cross-correlating with galaxy surveys helps to mitigate residual systematics from foregrounds, Radio Frequency Interference (RFI) and thermal noise \citep{Wolz:2015ckn,Pourtsidou:2016dzn}. Moreover, it can improve constraints on cosmological parameters and provide insight into the \hi\ astrophysics of the correlated galaxies \citep{Anderson:2017ert,Wolz:2021ofa}. \editt{A number of forthcoming telescopes are aiming to conduct \hi\ intensity mapping surveys, such as CHIME \citep{Newburgh:2014toa}, uGMRT \citep{Chakraborty:2020zmx}, Tianlai \citep{Li:2020ast}, HIRAX \citep{Newburgh:2016mwi}, CHORD \citep{Vanderlinde:2019tjt} and PUMA \citep{PUMA:2019jwd}, all of which are interferometers. There are also individual single-dish receivers such as FAST \citep{Bigot-Sazy:2015tot} and BINGO \citep{Wuensche:2018alk}. Currently, the \hi\ intensity mapping signal has only been detected in cross-correlation with galaxy surveys \citep{Chang:2010jp,Masui:2012zc,Anderson:2017ert,Li:2020pre,Tramonte:2020csa,Wolz:2021ofa,CHIME:2022kvg}.}

Both MeerKAT and the future SKA Observatory (SKAO) have been put forward as state-of-the-art intensity mapping instruments capable of complementing and extending cosmological measurements at other wavelengths \citep{Bacon:2018dui, Santos:2017qgq}. Using the single-dish data from each element of the array \citep{Battye:2012tg,Bull:2014rha}, we can access the large cosmological scales inaccessible by the interferometer due to its lack of very short baselines.

\editt{In this article, we used data from a MeerKAT pilot survey} to measure the cross-correlation power spectrum between the \hi\ signal and overlapping WiggleZ data \citep{Drinkwater:2009sd}. With only $10.5\,\text{hrs}$ of data for each of the 64 dishes over an effective survey area of ${\sim}\,200\,\text{deg}^2$, this detection shows the power of this approach and paves the way towards probing large cosmological scales with much larger surveys \edit{with} MeerKAT and SKAO.

The paper is structured as follows; in \secref{sec:ObsData} we introduce the data products used in this study. The formalism adopted for the power spectrum estimation and modelling is discussed in \secref{sec:Pkest&Mod}. \secref{sec:FGcleaning} introduces our approach to foreground cleaning in the MeerKAT intensity maps. We present our main results in \secref{sec:Results} and finally conclude in \secref{sec:Conclusion}.

\section{\edit{MeerKAT pilot survey data}}\label{sec:ObsData}


\edit{Here we summarise the MeerKAT instrument along with the observation strategy and pipeline used for obtaining the intensity mapping pilot survey data.} \edittt{The observational details are presented in \citep{Wang:2020lkn}} \editt{(hereafter W21) which describes in detail the calibration strategy and first sky maps analysis.}

\edit{MeerKAT is based in the Upper Karoo region of South  Africa and will eventually become a part of the final SKA1-Mid\footnote{\href{https://www.skao.int/}{www.skao.int}}. The SKAO is expected to end construction in July 2029, meaning MeerKAT will be used up to then to deliver transformative science \citep{Santos:2017qgq}. MeerKAT is composed of 64 dishes, each of which is $13.5\,\text{m}$ in diameter and can use three possible receivers, UHF-band $(580-1015\,\text{MHz})$, L-band $(900-1670\,\text{MHz})$ and S-band $(1750-3500\,\text{MHz})$. For the single-dish \hi\ intensity mapping pilot survey data in this work the L-band receiver was used, in principle allowing $z\,{<}\,0.58$ redshifts to be probed, however we only use a small sub-set of this range as discussed later in this section.}

\editt{For single-dish observations, we require a scanning strategy where the dishes are moved rapidly across the sky, covering a certain target patch.} \edit{The survey targeted a single patch of ${\sim}\,200\,\text{deg}^2$ in the WiggleZ 11hr field , covering $153^\circ\,{<}\,\text{R.A.}\,{<}\,172^\circ$ and $-1^\circ\,{<}\,\text{Dec.}\,{<}\,8^\circ$ \citep{Drinkwater:2009sd, Blake:2010xz,WiggleZ:2018def}, avoiding the strongest region of galactic emission. The observations took place over six nights between February and July 2019, which allowed all observing to be done at high elevation $({>}\,40\,\text{deg})$ to minimise fluctuations of ground spill and airmass.} \editt{After accounting for calibration in the observations, each complete scan across the sky patch took around 1.5 hours and we repeated this seven times over the six nights to give the 10.5 hours of combined data. We refer to each 1.5 hour scan as a \textit{time block}.} \edit{The 10.5 hours of observational data is obtained for each of the 64 dishes, although data from ${\sim}\,4$ dishes on average were completely removed from each} \editt{time} \edit{block due to equipment issues. The telescope scan speed ($5\,\text{arcmin}\,\text{s}^{-1}$ along azimuth) meant the dishes moved rapidly across the sky allowing $10\,\text{deg}$ to be scanned in ${\sim}\,100\,\text{s}$.}


\edit{Our calibration strategy for the time-ordered data (TOD) involved the use of noise diodes which were fired every $20$ seconds. Bandpass and absolute calibration of the diodes into Kelvin are performed through observation of a bright source of known flux density and spectrum.} \editt{We use 3C 273 as a bright source calibrator for five of the seven time blocks, using 3C 237 and Pictor A for the remaining blocks.} \edit{These diode solutions are then used to calibrate the TOD together with a multi-component model.} The use of the diode removes long-term noise correlation, so-called $1/\text{f}$ noise, due to receiver chain gain variations on time scales longer than 20\,s. On shorter time scales, $1/\text{f}$ noise is negligible compared to thermal noise fluctuations \citep{Li:2020bcr}. We also subtract the average signal every 220\,s in the TOD which suppresses \edit{residual very long timescale} gain changes. This should reduce the overall variance of the signal but can potentially have the adverse effect of removing \hi\ signal over large angular scales. In this work we assume any signal loss from this process is sub-dominant relative to the signal loss from foreground cleaning and thus do not attempt any reconstruction.

\edit{The data underwent three levels of RFI flagging at different stages of the calibration, starting with strong RFI flagging on the raw signal using the SEEK package \citep{AKERET20178}, then removing per-channel outliers in the TOD, then finally removing residual low-level RFI features, later in the pipeline after the map-making. Since these were dual-polarisation auto-correlation observations, each dish and time block provide separate HH and VV polarisation data at each frequency in the L-band. The mean-value of the two calibrated polarisation temperatures gives the total intensity corresponding to the Stokes I. At each frequency and time block,} the TOD for each dish, $d$, is projected into the map space via the map-making process \citep{Tegmark:1996qs},
\begin{align}
    \hat{m} = \left(A^\text{T} N^{-1} A\right)^{-1} A^\text{T} N^{-1} d
\end{align}
in which $A$ is the pointing matrix mapping the TOD to the map coordinates and $N$ is the noise covariance matrix between time stamps. 

The noise covariance matrix $N$ is assumed to be diagonal with constant variance during the observation, but we allow the variance to differ between dishes. The noise covariance is also projected to the map space via,
\begin{align}
    \hat{n} = \left(A^\text{T} N^{-1} A\right)^{-1},
\end{align}
where $\hat{n}$ is the pixel noise variance. The inverse of the pixel noise variance, $w_\hi\,{=}\,1/\hat{n}$, is used as the inverse-variance weight in the analysis\footnote{We adopt the \hi\ in the subscript of $w_\hi$ to maintain a consistent notation with later formalism where \edittt{the intensity maps weights} require distinguishing from the galaxy weights.}. We use the flat-sky approximation and grid the maps into square pixels with a width of $0.25\,\text{deg}$. To create the final maps, we average over all individual dish maps \edit{and time blocks at each frequency}.

\begin{figure}
    \centering
    \includegraphics[width=1\linewidth]{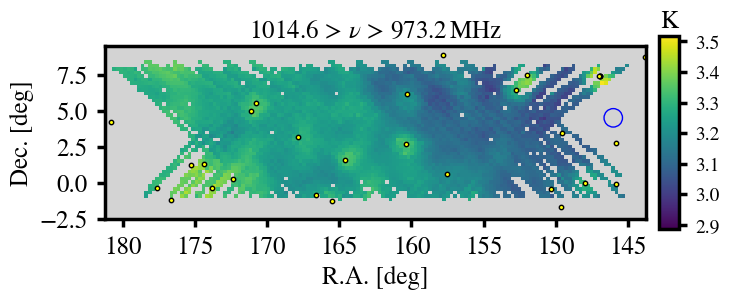}
    \caption{\edit{MeerKAT intensity map showing the total sky in the WiggleZ 11hr field, which is dominated by foreground emission. The map has been averaged over the 199 frequency channels covering $1014.6\,{>}\,\nu\,{>}\,973.2\,\text{MHz}$ ($0.400\,{<}\,z\,{<}\,0.459$). The yellow dots show the position of point sources with flux ${>}\,1\,\text{Jy}$ at $1400\,\text{MHz}$. The blue ring shows the FWHM of the MeerKAT beam at the mean frequency.}}
    \label{fig:FGmap}
\end{figure}

The MeerKAT L-band has 4096 frequency channels but in this work, we only use 199 channels at $973.2-1014.6\,\text{MHz}$ ($0.400\,{<}\,z\,{<}\,0.459$). \edit{Whilst using more frequency channels would in principle improve signal-to-noise in the cross-correlation, since there are WiggleZ galaxies available beyond these redshifts, we find the data to be more RFI dominated outside these frequency ranges, and flagging them is the safest option initially. Of the 199 channels selected to use}, a further 32 are removed due to their dominant contributions to the eigenmodes of the principal component analysis (see later discussion in \secref{sec:FGcleaning}) for those particular channels. This aggressive strategy can be seen as a final RFI-flagging stage and was necessary in order for us to obtain a cross-correlation detection. \edit{We leave to future work an investigation into a less aggressive flagging approach to see if this is beneficial in more sophisticated foreground cleaning algorithms. The observed map showing the total sky dominated by the foregrounds is shown in \autoref{fig:FGmap}. This shows good agreement with models of diffuse Galactic emission (see comparison in \editt{W21}) and a correlation can be seen by eye between the observed emission and the position of known point sources \editt{(obtained from the NED\footnote{\hyperlink{https://ned.ipac.caltech.edu/byparams}{ned.ipac.caltech.edu/byparams}} dataset)}, indicated by the yellow dots.}

\edit{The MeerKAT dish-diameter of $13.5\,\text{m}$ creates a large beam when used in single-dish mode. The full-width-half-maximum (FWHM) of the central lobe for the beam can be given approximately by $\theta_\text{FWHM}\,{=}\,1.16\,c\nu/D_\text{dish}$ \citep{Matshawule:2020fjz}, where $c$ is the speed of light, $\nu$ is the frequency of observation and $D_\text{dish}$ is the dish diameter. For the MeerKAT pilot survey this gives $\theta_\text{FWHM}\,{=}\,1.48\,\text{deg}$ at the mean frequency. This size of the beam is demonstrated in \autoref{fig:FGmap} by the blue ring which shows good agreement with the approximate sizes of the strongest point sources.}

\section{Power spectrum estimation \& modelling}\label{sec:Pkest&Mod}

\edit{Here we present the adopted formalism for the power spectrum estimation, along with the theoretical prediction used for comparison with the measurements. The power spectrum estimation process is based on the optimal weighting method outlined in \cite{Feldman:1993ky} (see \cite{Wolz:2015lwa,Blake:2019ddd} for applications to \hi\ intensity mapping cross-correlations). We also discuss our approach to error estimation. The theoretical model fitted to the measured power spectrum is similar to previous studies (e.g. \citet{Masui:2012zc,Wolz:2021ofa}) whereby we use a biased matter power spectrum with a fixed cosmology, with some treatment to account for linear redshift-space distortions (RSD) and damping from the telescope beam.}

\subsection{\edit{Power spectrum estimation}}\label{sec:Pkest}

We begin by defining the two Fourier transformed fields of the \hi\ temperature fluctuation maps $\delta T_\hi$ and the galaxy count field $n_\text{g}$ as
\begin{equation}
    \tilde{F}_\hi(\mathbf{k})=\sum_{\mathbf{x}} \delta T_\hi(\mathbf{x}) w_\hi(\mathbf{x}) \exp (i \mathbf{k}{\cdot}\mathbf{x})
\end{equation}
\begin{equation}
    \tilde{F}_\text{g}(\mathbf{k})=\sum_{\mathbf{x}} n_\text{g}(\mathbf{x}) w_\text{g}(\mathbf{x}) \exp (i \mathbf{k}{\cdot}\mathbf{x}) - N_\text{g}\tilde{W}_\text{g}(\mathbf{k})\,,
\end{equation}
where $N_\text{g}\,{=}\,\sum n_\text{g}$ is the total number of galaxies in the optical maps and $\tilde{W}_\text{g}$ is the weighted Fourier transform of the selection function ($W_\text{g}$)
which is normalised such that $\sum_\mathbf{x}W_\text{g}(\mathbf{x})\,{=}\,1$;
\begin{equation}
    \tilde{W}_\text{g}(\mathbf{k})=\sum_{\mathbf{x}} W_\text{g}(\mathbf{x}) w_\text{g}(\mathbf{x}) \exp (i \mathbf{k}{\cdot}\mathbf{x})\,.
\end{equation}
The selection function $W_\text{g}$ accounts for incompleteness in the WiggleZ survey and is constructed by stacking the random catalogues generated in \cite{Blake:2010xz}, reproducing the varying target and redshift completeness. The map weights in the above equations are the inverse variance map for the \hi\ field, $w_\hi(\mathbf{x})\,{=}\,1/\hat{n}(\mathbf{x})$ (defined in \secref{sec:ObsData}) and for the galaxies we use the optimal weighting as per \cite{Feldman:1993ky}
\begin{equation}\label{eq:FKPweight}
    w_\text{g}(\mathbf{x}) = 1\bigg/\left(1+\frac{W_\text{g}(\mathbf{x})N_\text{g}P_0}{V_\text{cell}}\right)\,,
\end{equation}
\begin{equation}
    \small
    \hat{P}_{\hi,\text{g}}(\mathbf{k}) = \frac{V_\text{cell}}{\sum\limits_\mathbf{x} w_\hi(\mathbf{x})w_\text{g}(\mathbf{x})W_\text{g}(\mathbf{x})}\operatorname{Re}\left\{\tilde{F}_\hi(\mathbf{k}){\cdot} \tilde{F}^{*}_\text{g}(\mathbf{k})\right\}\frac{1}{N_\text{g}}\,.
\end{equation}
Similarly, the \hi\ and galaxy auto power spectra, which are needed for the error estimation (see below), are given by
\begin{equation}\label{eq:HIautoPk}
    \hat{P}_\hi(\mathbf{k}) = \frac{V_\text{cell}}{\sum_\mathbf{x} w_\hi^2(\mathbf{x})}\lvert\tilde{F}_\hi(\mathbf{k})\rvert^2\,,
\end{equation}
\begin{equation}
    \hat{P}_\text{g}(k) = \frac{V_\text{cell}}{\sum_\mathbf{x} w^2_\text{g}(\mathbf{x})W^2_\text{g}(\mathbf{x})} \left[|\tilde{F}_\text{g}(\mathbf{k})|^2 - S_\text{g}\right]\frac{1}{N^2_\text{g}}\,,
\end{equation}
where $S_\text{g}$ accounts for the shot noise in the galaxy survey, given as
\begin{equation}
    S_\text{g} = N_\text{g}\sum_\mathbf{x} w_\text{g}^2(\mathbf{x})W_\text{g}(\mathbf{x})\,.
\end{equation}
These power spectra are all spherically averaged into bandpowers $\lvert\mathbf{k}\rvert\,{\equiv}\,k$ to provide the final 1D power spectra results
. For noise-dominated \hi\ intensity maps, the errors for the cross power can be estimated analytically from
\begin{equation}\label{eq:Pkerr}
    \small
    \hat{\sigma}_{\hi,\text{g}}(k)=\frac{1}{\sqrt{2 N_\text{m}(k)}} \sqrt{\hat{P}^2_{\hi,\text{g}}(k)+\hat{P}_\hi(k)\left(\hat{P}_\text{g}(k)+\frac{1}{\bar{n}_\text{g}}\right)}\,,
\end{equation}
where $N_\text{m}$ is the number of modes in each $k$-bin and $\bar{n}_\text{g}\,{=}\,N_\text{g}/(l_\text{x}\,{\times}\,l_\text{y}\,{\times}\,l_\text{z})$ is the number density of galaxies. The $1/\sqrt{2}$ factor in \autoref{eq:Pkerr} appears because this is the error on a cross-correlation of two fields, so the number of independent pairs available to measure the variance on the mean doubles. We compared these analytical error estimations to ones calculated from cross-correlating the MeerKAT data with the random WiggleZ catalogues used to derive the
selection function, finding very good agreement across all scales. Furthermore, we also found good agreement with an internal error estimation which used a jackknife approach \citep{Norberg:2008tg}.

\subsection{Theoretical modelling}\label{sec:Model}

We fit a model to the cross power spectrum which is given by
\begin{multline}\label{eq:Pkmodel}
    P_{\hi,\text{g}}(\mathbf{k}) = \overline{T}_\hi b_\hi b_\text{g} r (1+f\mu^2)^2 \,P_\text{m}(k) \\ \times\, \exp\left[\frac{-(1-\mu^2)k^2 R_\text{beam}^2}{2}\right]\,,
\end{multline}
where $\overline{T}_\hi$ is the mean \hi\ temperature of the field in mK, $b_\hi$ and $b_\text{g}$ are the \hi\ and galaxy biases and $r$ is the cross-correlation coefficient. We account for linear redshift-space distortions (RSD) with the $(1+f\mu^2)^2$ factor \citep{Kaiser:1987qv}, where $f$ is the growth rate of structure and $\mu$ is the cosine of the angle from the line-of-sight. $P_\text{m}$ is the matter power spectrum produced using \texttt{CAMB} \citep{Lewis:1999bs} with a Planck18 \citep{Planck:2018nkj} cosmology. The exponential factor approximates the smoothing of perpendicular modes due to the MeerKAT beam, where $R_\text{beam}$ is the standard deviation of the Gaussian beam profile in comoving units, taking into account the reconvolution \edit{(as later explained in \secref{sec:Resmooth})}, which gives $R_\text{beam}\,{=}\,13.3\,\text{Mpc}\,h^{-1}$. 

To fully account for RSD, the model in \autoref{eq:Pkmodel} should be (not including the beam damping for brevity)
\begin{equation}
    P_{\hi ,\text{g}}(\mathbf{k}) = \overline{T}_\hi\left[r b_\hi b_\text{g} + b_\hi f \mu^2 + b_\text{g} f \mu^2 + f^2 \mu^4\right] P_\text{m}(k)\,,
\end{equation}
which accounts for the biases appearing in cross-terms from the expansion of the two fields in redshift space $\delta^\text{s}_\hi(\mathbf{k})\,{=}\,b_\hi \delta_\text{m}(\mathbf{k})\,{+}\,f\mu\,\theta(\mathbf{k})$ and $\delta^\text{s}_\text{g}(\mathbf{k})\,{=}\,b_\text{g} \delta_\text{m}(\mathbf{k})\,{+}\,f\mu\,\theta(\mathbf{k})$, where $\theta(\mathbf{k})$ is the velocity divergence field\footnote{The cross-correlation coefficient $r$ only enters on cross-correlation between biased density terms.}. However, we are only attempting a fit to the spherically-averaged power spectrum monopole, which is uniformly averaged across $\mu$. This would make $b_\hi$ perfectly degenerate with $\overline{T}_\hi r$. To break this degeneracy we would need to introduce an anisotropic sensitivity on $\mu$ in our analysis, achieved by modelling the quadrupole \citep{Cunnington:2020mnn,Soares:2020zaq}. This would require a higher $S/N$ than we have available from the MeerKAT pilot survey data. This is why we follow previous literature \citep{Wolz:2021ofa} \editt{(hereafter W22)} and probe the degenerate quantity $\overline{T}_\hi b_\hi r$, but include a matter-only RSD, to avoid biasing the amplitude of the power spectrum by the $(1+ f\mu^2)\,{\sim}\,1.7$ Kaiser term.

The model in \autoref{eq:Pkmodel} is discretised onto the same 3D grid of modes as the data and then convolved with the survey window functions
\begin{multline}\label{eq:ConvolvedModel}
    P_{\hi,\text{g}}(\mathbf{k}) \rightarrow P_{\hi,\text{g}} *  W_\hi W_\text{g} = \\ \frac{\sum_{i}  P_{\hi,\text{g}}(\mathbf{k}_{i}^{\prime}) \operatorname{Re}\left\{\tilde{W}_\text{g}(\mathbf{k}-\mathbf{k}_{i}^{\prime}) \tilde{W}_\hi(\mathbf{k}-\mathbf{k}_{i}^{\prime})^{*}\right\}}{\sum w_\text{g}(\mathbf{x}) w_\hi(\mathbf{x}) W_\text{g}(\mathbf{x}) W_\hi(\mathbf{x})}\,.
\end{multline}
In lieu of a precisely constructed survey selection function $W_\hi$ for the \hi\ intensity maps, we use a simple binary window function that is 1 wherever a pixel is filled and 0 otherwise. The convolved model in \autoref{eq:ConvolvedModel} is spherically averaged into the same $k$-bins as the data.

The mean \hi\ temperature $\overline{T}_\hi$ can be recast to the \hi\ density fraction ($\Omega_\hi$) using \citep{Battye:2012tg}
\begin{equation}\label{eq:TbarModelEq}
    \overline{T}_\hi(z) = 180\,\Omega_{\hi}(z)\,h\,\frac{(1+z)^2}{\sqrt{\Omega_\text{m}(1+z)^3 + \Omega_\Lambda}} \, {\text{mK}} \,,
\end{equation}
where $\Omega_\text{m}$ and $\Omega_\Lambda$ are the density fractions for matter and the cosmological constant, respectively. Thus fitting the amplitude of the cross power spectrum allows us to constrain $\Omega_\hi b_\hi r$. When fitting $\Omega_\hi b_\hi r$ to the power spectrum data using the model in \autoref{eq:Pkmodel}, we fix the galaxy bias ($b_\text{g}$) and growth rate ($f$), since they are well constrained from other experiments relative to the other parameters. We assume $f\,{=}\,0.737$ (based on $f\,{\sim}\,\Omega_\text{m}(z)^\gamma$, where $\gamma\,{=}\,0.545$ \citep{Linder:2005in,Planck:2018nkj}) and $b_\text{g}\,{=}\,0.911$ \citep{Blake:2011rj} at the central redshift of our data ($z_\text{eff}\,{=}\,0.43$).

\section{Foreground Cleaning}\label{sec:FGcleaning}

Here we discuss the foreground cleaning performed on the MeerKAT intensity maps. We provide detailed descriptions on each stage in the following sub-sections but begin with a summary of the foreground cleaning method we adopt.

Before cleaning, the maps are resmoothed using a Gaussian window function with kernel size $1.2$ times the largest beam size within the frequency range (see \secref{sec:Resmooth} for details). The foreground cleaning is then performed using a blind Principal Component Analysis (PCA) method, which relies on the foregrounds being the dominant signal and correlated in frequency. Thus, by removing the first $N_\text{fg}$ principal component modes in frequency from each pixel, the majority of their contribution is suppressed (see \secref{sec:PCA} for further discussion). Foreground cleaning is imperfect, and the cleaned maps contain residual foreground. Furthermore, some \hi\ signal will be removed, typically on larger scales where modes are most degenerate with the spectrally smooth foregrounds. We aim to reconstruct this lost signal with a foreground transfer function, which we discuss in \secref{sec:TF}.

\subsection{Reconvolution of maps}\label{sec:Resmooth}

It is understood that a frequency-dependent beam size can cause the foregrounds to leak into a greater number of spectral modes, requiring more aggressive cleaning \citep{Switzer:2015ria,Alonso:2014dhk}. A way to potentially mitigate this issue is to convolve all maps to a common resolution before performing the foreground clean, as done in previous experiments \editt{\citep[W22]{Masui:2012zc,Wolz:2015lwa,Anderson:2017ert}}. However, recent tests on simulations suggest that a simple Gaussian resmoothing of the data to a common resolution does not improve blind foreground removal techniques, even if the true beam is a perfect Gaussian \citep{Matshawule:2020fjz,Spinelli:2021emp}. For real data though, it is beneficial to resmooth to homogenise some of the systematic contributions from e.g. residual RFI or polarisation leakage. For this reason we perform a weighted resmoothing on the MeerKAT \hi\ intensity maps prior to foreground cleaning.

An intensity map $\delta T^\prime$ which has a frequency dependent beam (denoted by the ${^\prime}$ index) with a FWHM $\theta_\text{FWHM}(\nu)$ in degrees, and an angular separation between pixels given by $\Delta\theta$, is convolved with the following kernel:
\begin{equation}\label{eq:ResmoothKernel}
    K(\Delta \theta,\nu)=\exp \left[-\frac{\Delta \theta^{2}}{2[\gamma \sigma_{\max }^{2}-\sigma^{2}(\nu)]}\right]\,,
\end{equation}
where $\sigma(\nu)=\theta_{\mathrm{FWHM}}(\nu) /(2 \sqrt{2 \ln 2})$, $\sigma_\text{max}$ is the maximum $\sigma(\nu)$ value
and $\gamma$ is a scaling factor which governs how much the final effective resolution is decreased by. 

In previous Green Bank Telescope (GBT) studies, a choice of $\gamma\,{=}\,1.4$ was used \citep{Masui:2012zc}. In this work, due to the already large MeerKAT beam, we use a smaller value of $\gamma\,{=}\,1.2$. \edit{We experimented with this choice, and discuss the consequences of varying $\gamma$ in our results (\secref{sec:Results}). The choice of $\gamma\,{=}\,1.2$ gives a frequency independent effective beam size of $\gamma\theta_\text{FWHM}(\nu_\text{min})\,{=}\,1.82\,\text{deg}$.}

The kernel in \autoref{eq:ResmoothKernel} is normalised such that the sum over all pixels is equal to unity, then the weighted convolution used to resmooth the maps is given by 
\begin{equation}
    \delta T(\theta,\nu) = \frac{\left[\delta T^\prime(\theta,\nu)\,w_\hi^\prime(\theta,\nu)\right] * K(\Delta\theta,\nu)}{w_\hi^\prime(\theta,\nu) * K(\Delta\theta,\nu)}\,,
\end{equation}
where $w_\hi^\prime(\theta)$ is the inverse variance weight. The $*$ denotes a convolution performed separately in each frequency channel e.g. $w_\hi^\prime(\theta) * K(\Delta\theta) = \Sigma_i w_\hi^\prime(\theta_i)K(\theta - \theta_i)$, \edit{where the summation is over each pixel $i$}. To ensure the weight field still represents the inverse variance of the new resmoothed field, the weight $w_\hi^\prime$ is convolved according to
\begin{equation}
    w_\hi(\theta,\nu) = \frac{\left[w_\hi^\prime(\theta,\nu) * K(\Delta\theta,\nu)\right]^2}{w_\hi^\prime(\theta,\nu) * K^2(\Delta\theta,\nu)}\,.
\end{equation}

\subsection{PCA foreground cleaning}\label{sec:PCA}

In this work, a PCA-based blind foreground subtraction method is used. The observed intensity maps can be represented by a matrix $\mathbf{X}_\text{obs}$ with dimensions $N_\nu\,{\times}\,N_\theta$ where $N_\nu$ is the number of frequency channels along the line-of sight and $N_\theta$ is the number of pixels. The assumption behind blind-foreground cleaning is that the data can be represented by the linear system $\mathbf{X}_\text{obs}\,{=}\,\mathbf{\hat{A}}\mathbf{S}\,{+}\,\mathbf{R}$, where $\mathbf{\hat{A}}$ represents the mixing matrix and $\mathbf{S}$ are the $N_\text{fg}$ separable source maps identified by projecting the mixing matrix along the data $\mathbf{S}\,{=}\,\mathbf{\hat{A}}^\text{T}\mathbf{X}_\text{obs}$. In PCA, the mixing matrix is extracted from the eigen-decomposition of the frequency covariance matrix of the mean-centred data, defined by $\mathbf{C}\,{=}\,(\mathbf{w}\mathbf{X}_\text{obs})^\text{T}(\mathbf{w}\mathbf{X}_\text{obs})/(N_\theta\,{-}\,1)$, where $\mathbf{w}$ are the inverse variance weights recast into $N_\nu\,{\times}\,N_\theta$ matrices. The eigen-decomposition is then given as $\mathbf{C}\mathbf{V}=\mathbf{V}\mathbf{\Lambda}$, where $\mathbf{\Lambda}$ is the diagonal matrix of eigenvalues ordered by descending magnitude, and $\mathbf{V}$ are the eigenvectors, the first $N_\text{fg}$ of which supplies the set of functions used to construct the mixing matrix. We assume the subtraction of $\mathbf{\hat{A}}\mathbf{S}$ in the linear system will remove dominant foregrounds, leaving behind in the residuals $\mathbf{R}$ \edit{most of the} \hi\ signal not removed in the subtraction \editt{along with Gaussian thermal noise}.

\begin{figure}
    \centering
    \includegraphics[width=1\linewidth]{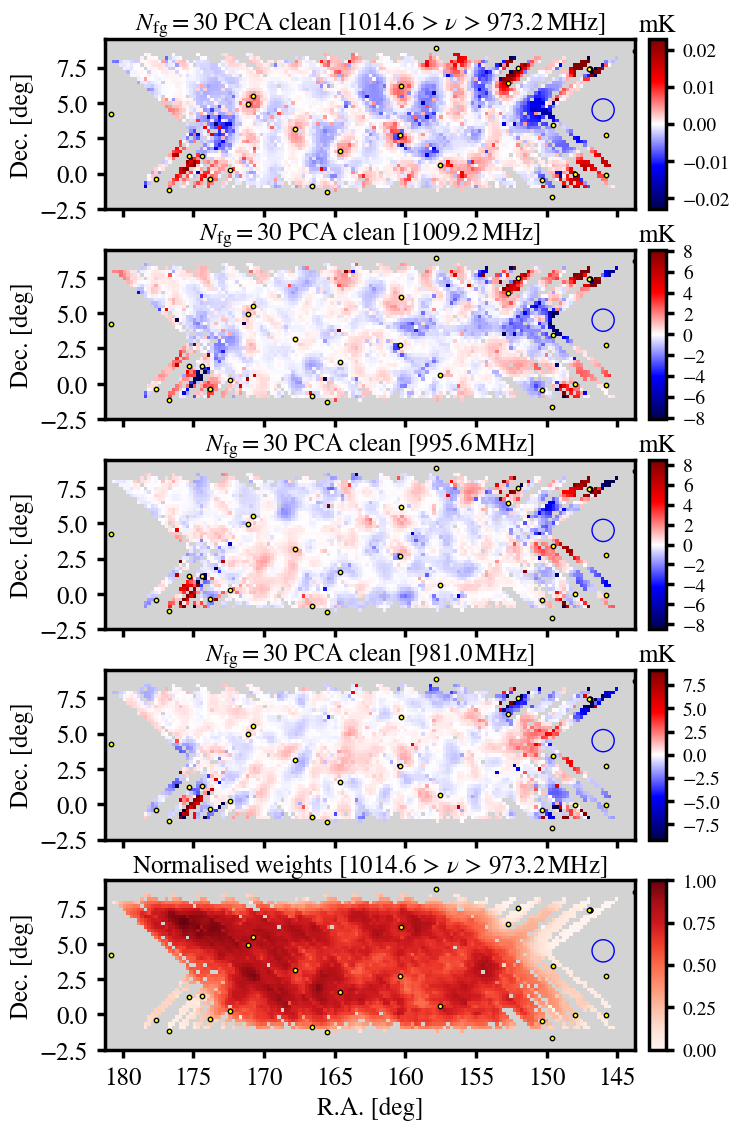}
    \caption{\edit{MeerKAT intensity maps foreground cleaned by removing $N_\text{fg}\,{=}\,30$ PCA modes. Top panel has been averaged over the 199 frequency channels covering $1014.6\,{>}\,\nu\,{>}\,973.2\,\text{MHz}$ ($0.400\,{<}\,z\,{<}\,0.459$) (foreground clean performed before averaging through frequency). Middle three panels show some example single frequency channels.   Bottom panel shows the normalised weights $w_\hi$ used in the analysis. The yellow dots show the position of point sources with flux ${>}\,1\,\text{Jy}$ at $1400\,\text{MHz}$. The blue ring shows the FWHM of the MeerKAT beam at the minimum frequency, multiplied by the resmoothing parameter i.e. $\gamma\theta_\text{FWHM}(\nu_\text{min})$.}}
    \label{fig:MKmaps}
\end{figure}

\edit{We show some resulting maps from the foreground cleaning in \autoref{fig:MKmaps}. The top panel, which shows the average through frequency, reveals evidence of residual foreground structure which can be seen from comparison with the uncleaned sky map in \autoref{fig:FGmap}. However, the amplitude of the map has decreased by several orders of magnitude, thus the foreground residuals should dominate less over the \hi\ fluctuations. The reason for the very low amplitude in the top panel is due to the average through frequency being suppressed in the PCA clean. Since this will remove large radial modes it can conceptually be seen as removing the mean from the line-of-sight. The middle three panels show some examples of cleaned maps in individual frequency channels. Here the amplitude is not as suppressed but there is less evidence of residual foreground structure and these maps are more likely dominated by frequency-varying systematics or residual RFI. This is more pronounced in the edges of the map which receive less observation time and are thus down-weighted, as shown by the bottom panel which presents the normalised weights, $w_\hi$, used in this analysis (see \secref{sec:ObsData}).}

\subsection{\edit{Foreground removal transfer function}}\label{sec:TF}

\begin{figure*}
    \centering
    \includegraphics[width=1\textwidth]{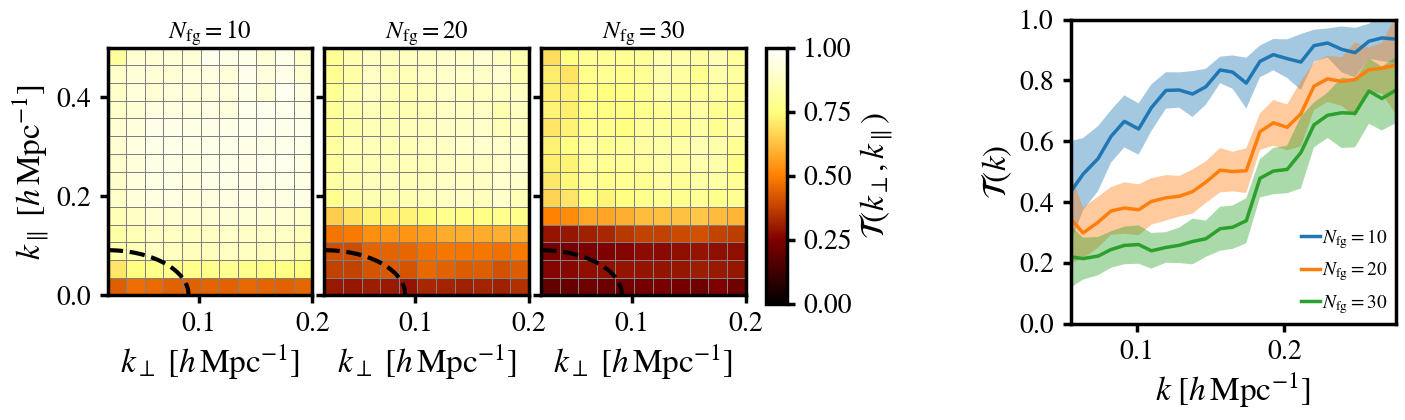}
    \caption{Foreground transfer functions used to correct for the signal loss in the power spectrum measurement from foreground cleaning. A value of $\mathcal{T}\,{=}\,1$ denotes no signal loss, while $\mathcal{T}\,{\ll}\,1$ denotes severe signal loss. Three-left panels show the transfer function decomposed into $k_\perp$ and $k_\parallel$ modes for different foreground cleans indicated by $N_\text{fg}$. The black-dashed line marks a characteristic scale of $\lvert k\rvert\,{=}\,0.08\,h\,\text{Mpc}^{-1}$ (approximately the scale of the first BAO wiggle maximum). Far-right panel shows the transfer function binned into the same spherically averaged $k$-bins as those used for the main power spectrum estimation. The shaded regions indicate the $1\sigma$ errors estimated from the variance in the 100 simulations used in the construction of $\mathcal{T}(k)$.}
    \label{fig:TF2D}
\end{figure*}

We compensate the signal loss due to the foreground cleaning with a transfer function. Following previous literature \citep{Masui:2012zc,Switzer:2013ewa,Switzer:2015ria}, the transfer function can be constructed by injecting mock intensity mapping data $\mathbf{M}_\hi$ into the true observed data $\mathbf{X}_\text{obs}$, which includes foregrounds and observational systematics. By running a PCA clean on this combination, we can measure (and compensate for) the signal loss in the cleaned mock data;
\begin{equation}\label{eq:MockforTF}
	\mathbf{M}_\text{c} = [\mathbf{M}_\hi + \mathbf{X}_\text{obs}]_\text{PCA} - [\mathbf{X}_\text{obs}]_\text{PCA}\,.
\end{equation}
The $[\ ]_\text{PCA}$ notation represents performing the PCA clean (outlined in \secref{sec:PCA}) on the quantities inside the brackets, treating them as a single combination. For example, the mixing matrix is not determined separately for both mock and data in $[\mathbf{M}_\hi + \mathbf{X}_\text{obs}]_\text{PCA}$, but determined for the combination of $\mathbf{M}_\hi + \mathbf{X}_\text{obs}$. We also subtract the PCA clean of the data, $[\mathbf{X}_\text{obs}]_\text{PCA}$, since this only adds uncorrelated variance, thus subtracting it makes convergence to a smooth transfer function more efficient, requiring fewer mock iterations. After calculating $\mathbf{M}_\text{c}$, we measure the cross power spectrum with a corresponding mock galaxy map $\mathbf{M}_\text{g}$, then divide this by a foreground-free equivalent to estimate the signal loss at each mode;
\begin{equation}\label{eq:TF}
	\mathcal{T}(k) = \left\langle  \frac{\mathcal{P}(\mathbf{M}_\text{c}\, ,\, \mathbf{M}_\text{g})}{\mathcal{P}(\mathbf{M}_\hi\, ,\, \mathbf{M}_\text{g})} \right\rangle \, .
\end{equation}
$\mathcal{P}()$ denotes an operator which measures the cross power spectrum, then spherically averages modes into the same $k$-bins as the data. The angled brackets represent an ensemble average over a large number of mocks (we use 100 in this work).

The \hi\ mocks are generated with the lognormal method \citep{ColesLognormal1991}, sampled from a model \hi\ power spectrum (the same as that used in the power spectrum fitting, see \secref{sec:Model}) with a Gaussian smoothing applied perpendicular to the line-of-sight to approximately emulate the MeerKAT beam. Similarly, for the galaxy mocks, we generate a lognormal density field with the same random seed as the \hi, then Poisson sample galaxies onto the field with the same number count as the WiggleZ catalogue. These steps ensure the mock fields emulate the amplitudes of real maps as realistically as possible. There is evidence suggesting the transfer function is not overly sensitive to the choice of fiducial cosmology \citep{Cunnington:2022ryj}, but further investigation into how much it can be relied on for precision cosmology is required.

\edit{We plot the Fourier-space transfer function $\mathcal{T}(k_\perp,k_\parallel)$ in \autoref{fig:TF2D} (3 left panels) decomposed into anisotropic $k$-bins perpendicular ($k_\perp\,{=}\,\sqrt{k^2_\text{x} + k^2_\text{y}}$) and parallel ($k_\parallel\,{\equiv}\,k_\text{z}$) to the line-of-sight, as well as $\mathcal{T}(k)$ (far-right panel). As the $\mathcal{T}\,{<}\,1$ values show, the foreground cleaning is causing signal loss, mostly in the small-$k_\parallel$ modes.}

Any power spectrum measurement we make on the data is divided by $\mathcal{T}(k)$ to correct for the signal loss (unless clearly stated in demonstrative figures). For the \hi\ auto power spectrum used in the error estimation (outlined in \secref{sec:Pkest}), we also multiply through by $1/\mathcal{T}(k)$. Previous studies have opted to use $1/\mathcal{T}(k)^2$ as a correction instead \citep{Switzer:2013ewa}, motivated by the assumption that in auto-correlation, signal loss occurs in both maps, so there should be twice the reconstruction of power needed. However, from simulation tests, we found this over-corrected the signal loss. Furthermore, our analytical error estimation on the cross power spectrum, which uses the auto-\hi\ power spectrum, is found to be in good agreement with other approaches of error estimation using the WiggleZ randoms and jackknife tests. $\chi^2_\text{dof}$ analysis also suggests our errors are not over-estimated in any case. This changes if we opt for the $1/\mathcal{T}(k)^2$ correction where it becomes clear that the errors have been over-estimated, suggesting that the signal loss in the auto-\hi\ power spectrum has been over-corrected. We defer further investigation into signal loss in the \hi\ auto-correlation to future work.

\section{Results}\label{sec:Results}

\edit{Here the main results are presented beginning with an analysis of the \hi\ auto-correlation power spectrum (\secref{sec:HIautoPk}) to provide some insight into the quality of the MeerKAT pilot intensity mapping data and the foreground cleaning performance. The main results from the cross-correlation with WiggleZ galaxies are then presented and analysed in \secref{sec:ResultsWigZcross} and lastly we provide some constraints on the \hi\  density parameter in \secref{sec:ResultsOmHI}.}

\subsection{\hi\ auto power spectrum}\label{sec:HIautoPk}

\begin{figure*}
    \centering
    \includegraphics[width=1\textwidth]{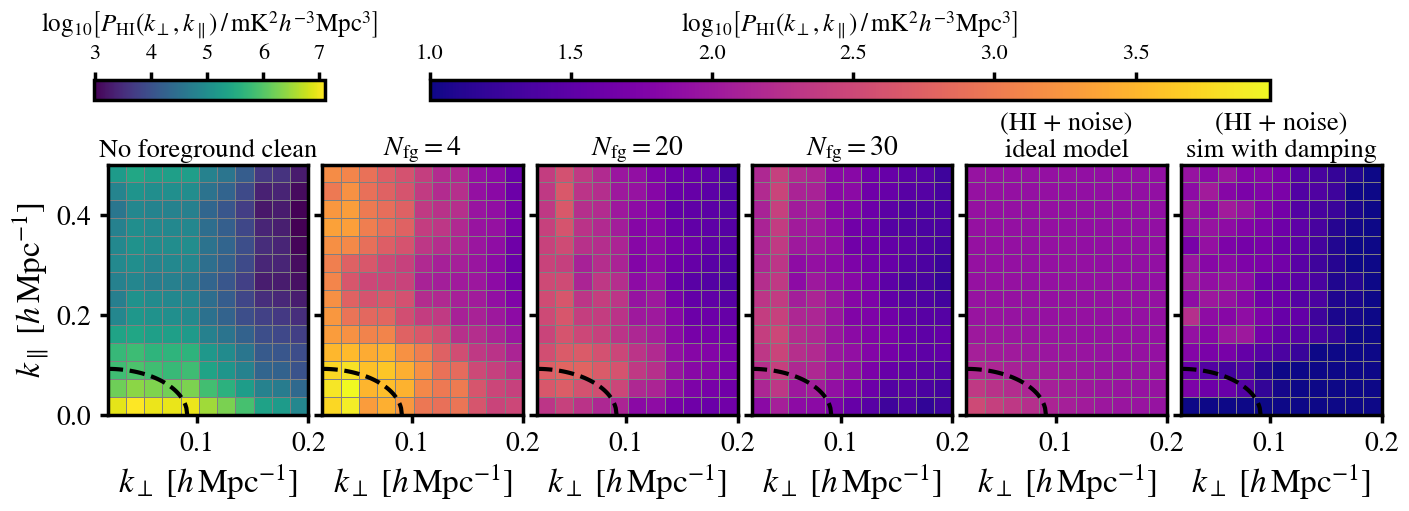}
    \caption{MeerKAT \hi\ auto power spectra at $0.400\,{<}\,z\,{<}\,0.459$ decomposed into $k_\perp$ and $k_\parallel$ modes for different foreground cleans indicated by $N_\text{fg}$ (the number of PCA modes removed). The fifth panel shows a predicted level for the \hi\ signal plus purely Gaussian thermal noise, in an ideal case assuming no signal loss. The thermal noise is predicted to be ${\sim}\,2\,\text{mK}$ for this survey (estimated in \editt{W21}). The far-right panel shows a simulation of \hi\ plus Gaussian noise including signal damping from the beam, additional resmoothing and foreground cleaning. As in \autoref{fig:TF2D}, the black-dashed line marks a characteristic scale of $\lvert k\rvert\,{=}\,0.08\,h\,\text{Mpc}^{-1}$ (approximately the scale of the first BAO wiggle maximum) at which, from our modelling, we would expect the \hi\ power to be $P_\hi\,{\sim}\,100\,\text{mK}^2h^{-3}\text{Mpc}^3$, i.e. $\log_{10}[P_\hi]\,{=}\,2$. None of these power spectra have had signal loss reconstructed by the transfer function.}
    \label{fig:HIAutoPk}
\end{figure*}

The auto power spectrum of the cleaned \hi\ intensity maps gives some indication of how much foregrounds have been suppressed. In \autoref{fig:HIAutoPk} we show the 2D auto-\hi\ power spectrum. The far-left panel shows the original data before foreground removal, demonstrating the dominance of the foregrounds and their concentration on the largest scales, particularly at small-$k_\parallel$. Removing just a few principal components reduces the amplitude of the auto power spectrum by several orders of magnitude, as shown in the second panel. In simulation tests \edit{using foreground models extrapolated from real data}, removing $N_\text{fg}\,{\sim}\,4$ PCA modes is sufficient to remove the majority of the foregrounds \citep{Alonso:2014dhk,Cunnington:2020njn}, and this should also be the case for a perfectly designed and calibrated experiment. The resulting cleaned maps in this idealised case would contain \hi\ signal and Gaussian thermal noise, similar to the model in the fifth panel of \autoref{fig:HIAutoPk}. The thermal noise in the ideal model is large due to the survey's low observing time. This yields an auto power spectrum amplitude of $P_\text{noise}\,{\sim}\,85\,\text{mK}^2h^{-3}\text{Mpc}^3$. \edit{The amplitude of the \hi\ signal varies with scale but should reach\footnote{We assume $\Omega_\hi b_\hi\,{=}\,0.85\,{\times}\,10^{-3}$ for the \hi\ model and simulation in \autoref{fig:HIAutoPk}.} $P_\hi\,{\sim}\,100\,\text{mK}^2h^{-3}\text{Mpc}^3$ at around $k\,{\sim}\,0.08\,h\text{Mpc}^{-1}$ and fall to $P_\hi\,{\sim}\,10\,\text{mK}^2h^{-3}\text{Mpc}^3$ at $k\,{\sim}\,0.18\,h\text{Mpc}^{-1}$, therefore we expect the noise to} dominate over the \hi\ signal on most scales, as seen in the \autoref{fig:HIAutoPk} model.

For this pilot-survey data, \editt{instrumental calibration imperfections, residual RFI and other systematics (see discussions in \cite{Irfan:2021xuk}) will distort the idealised model in \autoref{fig:HIAutoPk}}. The instrumental response modulates the foregrounds, resulting in additional spectral structure that requires more PCA modes to be removed. Uncleaned modes containing residual foregrounds, RFI and other systematics will positively bias the auto power spectrum\footnote{This assumes these systematics have not caused the gain to be systematically overestimated.}. This is why for the $N_\text{fg}\,{=}\,4$ case, \editt{the power spectrum does} not reach the level of the \hi\ signal plus noise.

As $N_\text{fg}$ increases in \autoref{fig:HIAutoPk}, the amplitude of the power drops, and this occurs more severely for large modes, particularly at small-$k_\parallel$. \edit{Note we do not correct for signal loss with the transfer function in any of the auto power spectra in \autoref{fig:HIAutoPk} to allow for a more detailed examination.} While it appears that \editt{the power spectrum is} reaching below the idealised \hi\ + noise model at high-$k_\perp$ and low-$k_\parallel$ in the $N_\text{fg}\,{\ge}\,20$ cases, \editt{a detailed comparison} would need to account for the effects of the map reconvolution and the foreground clean, both of which would damp the \hi\ + noise model further. In the far-right panel we show a simulated \hi\ mock with the same noise level as the idealised model, but include some observational effects. To emulate the beam and reconvolution, we smooth the simulation perpendicular to the line-of-sight. We also emulate signal loss from the foreground clean by projecting out modes based on the same PCA mixing matrix functions derived for the $N_\text{fg}\,{=}\,30$ data case. Comparison between the $N_\text{fg}\,{=}\,30$ and far-right panel suggests \editt{the measured auto power spectrum from the data has not reached this estimated level}, indicating that residual RFI, foregrounds, and other systematics are present in the data. \edit{As supported by the foreground transfer functions in \autoref{fig:TF2D}, we know there is signal loss from the foreground clean, hence we are not free to arbitrarily increase the aggressiveness of the foreground clean to further reduce residuals. Thus, a balance is required between reducing foreground residuals and limiting cosmological signal loss.}

Since the additive bias from non-cosmological residuals is unknown, it is difficult to compare auto-correlated data and model and conclude that a cosmological detection has been achieved. Cross-correlating with galaxy surveys avoids these additive biases, serving the motivation for this work. We leave a detailed study into the auto-\hi\ power for future work, where we will explore cross-correlating different sets of dishes or observational time blocks (a method adopted in GBT experiments \citep{Masui:2012zc}).

\subsection{Cross-correlation with WiggleZ galaxies}\label{sec:ResultsWigZcross}

\begin{figure}
    \centering
    \includegraphics[width=\linewidth]{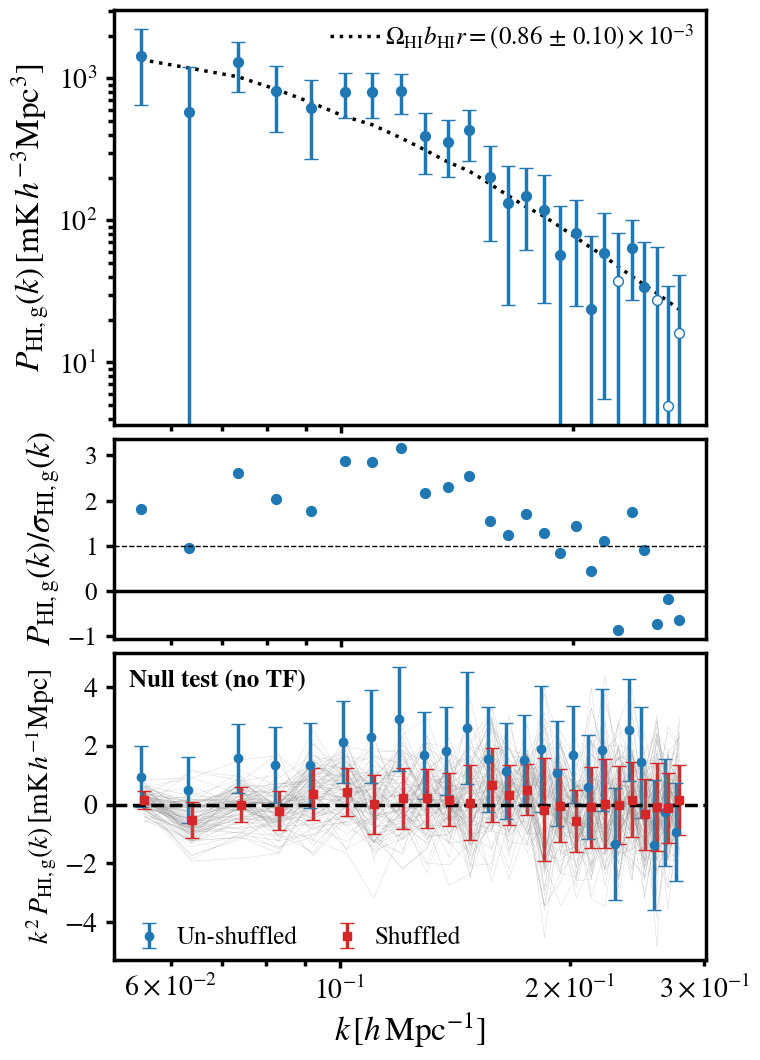}
    \caption{Cross power spectrum between WiggleZ galaxies and MeerKAT \hi\ intensity maps cleaned by removing $N_\text{fg}\,{=}\,30$ PCA modes at $0.400\,{<}\,z\,{<}\,0.459$, with $1\sigma$ error-bars (top panel). Hollow markers indicate a negative correlation. The black-dotted line represents a theoretical model (see \autoref{eq:Pkmodel}), fitted with an amplitude parameter $\Omega_\hi b_\hi r$ given in the top-right. Calculating the $\Delta\chi^2$ relative to a null-model ($P_{\hi,\text{g}}\,{=}\,0$) evaluates this as a $7.7\sigma$ cross-correlation detection. The middle panel shows the ratio between data and error. The bottom panel shows a null test where the WiggleZ galaxy maps have had been shuffled along redshift. The thin grey lines show 100 different shuffles. The average (red squares) and standard deviation (red error bars) across the shuffled samples are shown relative to the original (blue-dots). In both cases in the bottom panel, no scaling by the transfer function has been applied.}
    \label{fig:CrossPkSteve}
\end{figure}

There are 4031 galaxies in the overlapping 11hr field of the WiggleZ galaxy survey \citep{Drinkwater:2009sd,WiggleZ:2018def}. Following the steps outlined in \secref{sec:Pkest}, we compute an estimate for the cross power spectrum between the WiggleZ galaxies and the MeerKAT intensity maps, foreground cleaned by removing $N_\text{fg}\,{=}\,30$ PCA modes. We present this power spectrum in \autoref{fig:CrossPkSteve}. The middle panel shows the signal-to-noise ratio, where we find $S/N\,{\sim}\,2$ on large scales. We use an analytical method to estimate the errors (discussed in \secref{sec:Pkest}). At smaller scales, the MeerKAT beam \citep{Asad2021,deVilliers:2022vhg}, which is significantly larger than previous intensity mapping surveys\footnote{\edit{The FWHM for the central lobe of the MeerKAT beam is $1.82\,\text{deg}$ after resmoothing (see \secref{sec:ObsData} and \secref{sec:Resmooth}). For comparison, this is ${\sim}\,4$ times larger than the GBT ($100\,\text{m}$ dish diameter) observations at $z\,{\sim}\,0.8$, who had an effective resolution after resmoothing of $\gamma\theta_\text{FWHM}\,{\sim}\,0.44\,\text{deg}$ using $\gamma\,{=}\,1.4$ \citep{Masui:2012zc}.}} \citep{Masui:2012zc,Anderson:2017ert,Chakraborty:2020zmx,CHIME:2022kvg}, is the main reason for the poor $S/N$\edit{, since the signal is damped by the beam}.

The model (black-dotted line) in \autoref{fig:CrossPkSteve} is calculated following \secref{sec:Model}. In this model, we fix all parameters to fiducial quantities except for the degenerate quantity $\overline{T}_\hi b_\hi r$ for which we assume scale-independence, hence this quantity will only affect the amplitude of the model. Thus, by fitting the amplitude of the cross power spectrum, we are sensitive to $\Omega_\hi b_\hi r$ (from the relation in \autoref{eq:TbarModelEq}). We quote the best-fit value for $\Omega_\hi b_\hi r$ in the top-right of \autoref{fig:CrossPkSteve}, which was fitted to the data using a least-squares method. We discuss the parameter constraints on $\Omega_\hi$ in the following section.

We find a good agreement ($\chi^2_{\rm dof}\,{\sim}\,1$)\footnote{$\chi^2_{\rm dof}\,{\equiv}\,\chi^2/{\rm dof}$ where $\text{dof}\,{=}\,24$ are the degrees of freedom which is the number of $k$-bins minus 1 for the single parameter we fit.} between the data and model across all scales ($0.05\,{<}\,k\,{<}\,0.28\,h\,\text{Mpc}^{-1}$) in \autoref{fig:CrossPkSteve}. Furthermore, we analyse $\sqrt{\Delta \chi^2}\,{\equiv}\,\sqrt{\chi^2 - \chi^2_{\rm null}}$, the difference between the data's $\chi^2$ evaluated using our cross-correlation model, and one using a null model with zero cross power. This quantifies the statistical significance of the cross-correlation detection. We achieve a $7.7\sigma$ detection, providing strong evidence for the first verification of a cosmological signal with a multi-dish array performing single-dish intensity mapping.

We conducted various null tests on the analysis pipeline. The bottom panel of \autoref{fig:CrossPkSteve} shows the results from shuffling the galaxy maps along the line-of-sight, which should destroy the cross-correlation clustering signal. We re-ran the power spectrum estimation pipeline after each shuffle and found a result consistent with zero. We found similar null results when shuffling the cleaned \hi\ intensity maps along the line-of-sight and swapping the WiggleZ maps with random mocks from \cite{Blake:2010xz}. 

We did not apply the transfer function in the bottom panel of \autoref{fig:CrossPkSteve}, since scaling the null results would make no difference. \edit{Encouragingly, we were still able to obtain a detection for the original un-shuffled results where a transfer function has also not been used. This provides a ${>}\,4\sigma$ detection, which is strong evidence for correlated clustering since this result relied on no signal reconstruction. However, using the transfer function avoids biasing parameter estimates and improves the cross-correlation detection. We therefore implement it in the cross-correlation results.}

\edit{We experimented with the choice of the resmoothing parameter $\gamma$ (\secref{sec:Resmooth}), finding $\gamma\,{=}\,1$ produced a noisier power spectrum with worse model agreement
. The higher choice of $\gamma\,{=}\,1.4$ delivered a similarly good model agreement compared to $\gamma\,{=}\,1.2$, but had a slightly lower cross-correlation detection significance 
due to the increased damping at high-$k$. Not performing any reconvolution still delivered a clear detection but resulted in a particularly noisy power spectrum at small-$k$
, indicating the presence of residual foreground and systematics which are mitigated by this resmoothing procedure.}

\begin{figure}
    \centering
    \includegraphics[width=0.94\linewidth]{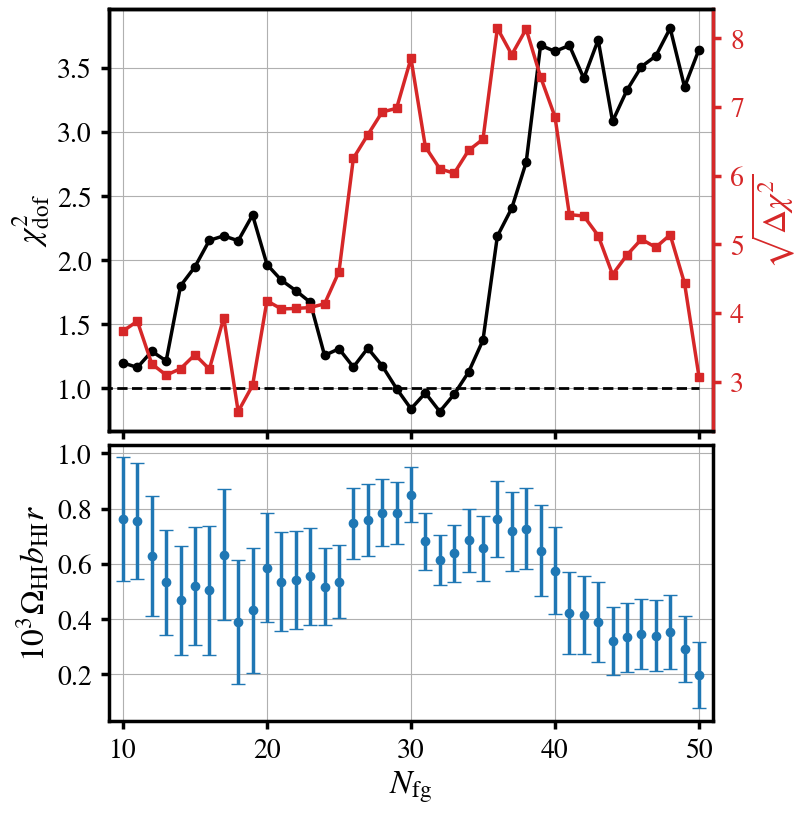}
    \caption{Sensitivity of results to foreground cleaning. Top panel, left-axis and black-dot data show the reduced $\chi^2$ for each foreground cleaned cross power spectra relative to its best-fit model. Top panel, right-axis and red-square data show the detection significance for the cross power spectra relative to a null model. Bottom panel shows the variation in the best-fit $\Omega_\hi b_\hi r$ for each $N_\text{fg}$ case, from a least-squares fit to the cross power spectrum amplitude.}
    \label{fig:Nfg_dependence}
\end{figure}

For the power spectrum in \autoref{fig:CrossPkSteve} we chose $N_\text{fg}\,{=}\,30$ as it provides an excellent goodness-of-fit ($\chi^2_\text{dof}$). \autoref{fig:Nfg_dependence} (top panel) shows how varying $N_\text{fg}$ influences $\chi^2_\text{dof}$, which should ideally be close to unity to represent a good model fit to the data with reasonably sized errors. For each $N_\text{fg}$ case we recalculate the transfer function and re-fit the free parameter $\Omega_\hi b_\hi r$ (values shown by the bottom panel), which avoids the $\chi^2_\text{dof}$ improving simply because the amplitude of the power is decreasing into agreement with a pre-selected fiducial $\Omega_\hi b_\hi r$. We also show the cross-correlation detection strength, given by $\sqrt{\Delta\chi^2}$, on the right-hand (red) axis of \autoref{fig:Nfg_dependence} (top panel).

At low $N_\text{fg}$, the $\chi^2_\text{dof}$ appears reasonable but this is due to the larger statistical errors on the cross power spectrum, which is fairly consistent with zero for these $N_\text{fg}$, as identified by the low detection significance in the $\sqrt{\Delta\chi^2}$ results. The errors are larger for low $N_\text{fg}$ because the residual foregrounds contribute significantly more variance to the maps, even though the residuals themselves are expected to correlate out on average. Increasing $N_\text{fg}$ from 10 to 20 does little to improve the detection significance and initially worsens the $\chi^2_\text{dof}$ caused by a decrease in error-bar size. At $N_\text{fg}\,{\sim}\,30$ enough components have been removed that a clear detection starts to manifest along with an improved agreement between data and model, given by the $\chi^2_\text{dof}\,{\sim}\,1$. Going to much higher $N_\text{fg}$ starts to over-clean the maps, reducing $S/N$ and worsening the detection. To justify this explanation for the deterioration in results for $N_\text{fg}\,{>}\,30$, we analysed the cross-correlation for maps constructed using just the principal components between 30 and 40. \edit{These maps} provided a ${\sim}\,3.4\sigma$ detection, indicating that a lot of signal is present in the modes with $30\,{<}\,N_\text{fg}\,{<}\,40$, which explains the deterioration in $\chi^2_\text{dof}$ beyond $N_\text{fg}\,{=}\,30$, where these modes are gradually removed. 

\edit{The main result we chose to present ($N_\text{fg}\,{=}\,30$) is picked from a region where $N_\text{fg}$ could be ${+/-}\,4$ of this choice and still deliver a ${>}\,6\,\sigma$ detection, thus representing robust evidence for cross-correlation. The choice of $N_\text{fg}\,{=}\,30$ offers a good balance between goodness-of-fit ($\chi^2_\text{dof}$) and detection significance ($\sqrt{\Delta\chi^2}$), as well as a compromise between reducing residual foregrounds and limiting signal loss, as discussed in \secref{sec:HIautoPk}. However, } \autoref{fig:Nfg_dependence} highlights the sensitivity of results to the foreground clean, and is further evidence that residual foregrounds and systematics are spread throughout the principal components. The ratio between systematics and signal varies among the various components thus some will be more influential on the cross-correlation than others. This causes a variation in the derived parameter $\Omega_\hi b_\hi r$, shown by the bottom panel of \autoref{fig:Nfg_dependence}. We estimate a contribution to the error budget of $\Omega_\hi b_\hi r$ caused by the variance across the different $N_\text{fg}$, discussed further in the following section. \edittt{An even more detailed understanding of the variation in results with $N_\text{fg}$, as well as signal loss correction uncertainties, must be gained from end-to-end simulations which we are pursuing for future work.}

\subsection{Constraints on \boldmath$\Omega_\hi$}\label{sec:ResultsOmHI}

Fitting the amplitude of the \hi-galaxy cross power spectrum provides a constraint on $\Omega_\hi b_\hi r$ as a function of redshift. From our results in \autoref{fig:CrossPkSteve}, we find $\Omega_\hi b_\hi r\,{=}\,[0.86\,{\pm}\,0.10\,({\rm stat})\,{\pm}\,0.12\,({\rm sys})]\,{\times}\,10^{-3}$. The systematic error accounts for uncertainty from the map calibration and variance in results from the choice of $N_\text{fg}$. Firstly, uncertainty from the map calibration could cause a bias to the overall amplitude of the power spectrum. We address this by studying the residuals relative to the model in our calibration study \editt{(W21)}. We are able to estimate that gain uncertainties should be at a level of ${\sim}\,2\%$. Secondly, we account for the variance in results from the different number of PCA modes removed, indicative of residual systematics. We do this by evaluating the standard deviation on all $\Omega_\hi b_\hi r$ fits (see \autoref{fig:Nfg_dependence}, bottom panel) from each reasonable choice of $N_\text{fg} \, (10\,{<}\,N_\text{fg}\,{<}\,40)$, which we find to be ${\sim}\,0.115$, equating to a ${\sim}\,13\%$ error on $\Omega_\hi b_\hi r$. The combination of these two error components added in quadrature yields the systematic error in our final constraint.

\begin{figure}
    \centering
    \includegraphics[width=\linewidth]{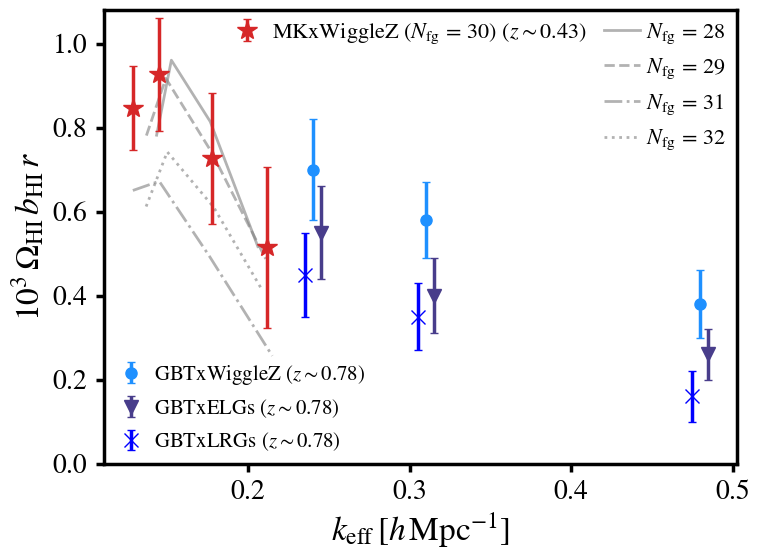}
    \caption{Constraints on $\Omega_\hi b_\hi r$ measured at different effective scales $k_\text{eff}$, achieved by \edit{removing} low-$k$ data points from the cross power spectrum, the data-points for each case are therefore not independent. The results for this paper are shown in red-stars using an $N_\text{fg}\,{=}\,30$ PCA clean and the four grey lines show results from different values of $N_\text{fg}$. For comparison we also plot the recent results from GBT cross-correlations \editt{(W22)}.}
    \vspace{0.2cm}
    \label{fig:OmHI}
\end{figure}

The \hi\ bias is not yet well understood but is expected to have some scale dependence when entering non-linear scales (high-$k$) \citep{Carucci:2015bra,Villaescusa-Navarro:2018vsg,Spinelli:2019smg}. Furthermore, the cross-correlation coefficient $r$, included to account for stochasticity between the two fields, will also have some scale dependence. We therefore examined how the constraint on $\Omega_\hi b_\hi r$ changed as we varied the scales at which it was measured. By \edit{removing} small-$k$ data points, we change the effective scale of the measurement, calculated by 
\begin{equation}\label{eq:keff}
    \edit{k_\text{eff} = \sum_i k_i (S/N)_i^2/\sum_i(S/N)_i^2\,,}
\end{equation}
where $(S/N)_i$ is the signal-to-noise ratio in each $k_i$ bin, i.e. $\hat{P}_{\hi,\text{g}}(k_i)/\hat{\sigma}_{\hi,\text{g}}(k_i)$. The scale dependence on the measurements of $\Omega_\hi b_\hi r$ is shown in \autoref{fig:OmHI} (red-star points) for $N_\text{fg}\,{=}\,30$. The other coloured data points show previous intensity mapping constraints from GBT cross-correlation with galaxy surveys at $z\,{\sim}\,0.8$ \editt{(W22)}. The GBT intensity maps had a significantly smaller beam than MeerKAT thus were able to probe higher-$k$. The MeerKAT and GBT measurements are at different redshifts so a direct comparison is not possible. Despite this, there still appears to be a trend with $k_\text{eff}$ suggesting a detection of scale dependence in $b_\hi r$. However, the continuity of the trend is affected by the choice of $N_\text{fg}$, as shown by the different grey lines in \autoref{fig:OmHI}. Furthermore, there is a possibility the scale dependence is influenced by systematics, which are mitigated in the scale cuts. The smallest-$k$ modes are the most affected by the transfer function, with up to 80\% increase in amplitude (see \autoref{fig:TF2D}). Thus, further investigation is needed to disentangle the scale-dependence of $b_\hi r$ from possible scale-dependent systematics. \edit{Despite using the same scale cuts in \autoref{fig:OmHI} for each $N_\text{fg}$ case, the effective scale is recalculated each time according to \autoref{eq:keff} and can therefore provide different $k_\text{eff}$ which explains} \editt{the slight horizontal offsets in the curves for different $N_\text{fg}$}.

By making some further assumptions on $b_\hi$ and $r$, we can isolate the constraint on $\Omega_\hi$. \edit{The cross-correlation coefficient $r$ is not currently well understood with more observational and simulation-based studies required. It will also vary depending on the \hi\ content of the selected optical galaxies \citep{Wolz:2015ckn}. However, on large scale it is reasonable to assume $r$ should be close to unity and reasonably independent of redshift. For our purposes, we therefore} assume $r\,{=}\,0.9\,{\pm}\,0.1$ \citep{Khandai:2010hs} and for the bias we interpolate between the hydrodynamic simulations in \cite{Villaescusa-Navarro:2018vsg} and use $b_\hi\,{=}\,1.13\,{\pm}\,0.10$. Fitting the power spectrum data in \autoref{fig:CrossPkSteve} across all scales ($k_\text{eff}\,{\sim}\,0.13\,h\,\text{Mpc}^{-1}$), provides a constraint of $\Omega_\hi\,{=}\,[0.83\,{\pm}\,0.15\,({\rm stat})\,{\pm}\,0.11\,({\rm sys})]\,{\times}\,10^{-3}$ at the redshift of $z\,{=}\,0.43$, which is reasonably consistent with other results in the literature (see comparison in \editt{W22}).

\section{Conclusion}\label{sec:Conclusion}

\hi\ intensity mapping is a novel method for probing large scale cosmic structure and will be a primary objective for the future SKAO. To achieve this, it is necessary for the multi-dish array to operate in single-dish (auto-correlation) mode, as opposed to a conventional interferometer. In this work we have demonstrated, for the first time, the successful detection of cosmological signal using the MeerKAT multi-dish array in single-dish mode. This represents a major milestone in demonstrating the feasibility of this survey mode for SKAO.

We achieved this by cross-correlating $10.5\,\text{hrs}$ of MeerKAT pilot survey intensity maps with overlapping optical galaxies from the  WiggleZ Dark Energy Survey. A measurement of the cross power spectrum between these fields provided a $7.7\sigma$ detection of a cross-correlation. We relied on an aggressive filtering process, removing 30 modes in a PCA-based foreground clean, necessary due to the presence of systematic contributions in the pilot survey data. This allowed us to obtain a constraint of $\Omega_\hi b_\hi r\,{=}\,[0.86\,{\pm}\,0.10\,({\rm stat})\,{\pm}\,0.12\,({\rm sys})]\,{\times}\,10^{-3}$ from fitting the amplitude of the cross power spectrum at an effective scale of $k_\text{eff}\,{\sim}\,0.13\,h\,\text{Mpc}^{-1}$. Varying the effective scale of the measurement changed the value for $\Omega_\hi b_\hi r$, something noted in previous studies \editt{(W22)}. We also found $\Omega_\hi b_\hi r$ to have a dependence on the number of foreground modes removed, so we included this variance in the systematic error budget of the constraint. The ${\sim}\,17.8\%$ precision represents a competitive $\Omega_\hi b_\hi r$ constraint relative to other intensity mapping experiments. Furthermore, with additional assumptions on $b_\hi$ and $r$, we provided insight into the cosmic \hi\ density $\Omega_\hi$, for which measurements at higher redshifts are vital for understanding the evolution of \hi.

The MeerKAT telescope will continue to conduct \hi\ intensity mapping observations in single-dish mode. With enhanced calibration techniques and more observing time, improved constraints will be possible with less aggressive foreground removal. With this we can attempt a detection of the \hi\ in auto-correlation, which is yet to be achieved. Observations have now been conducted in MeerKAT's UHF band ($0.40\,{<}\,z\,{<}\,1.45$), opening the possibility of higher redshift probes and for cross-correlating UHF-band data with the L-band data used in this work, with the aim of mitigating systematics.

\section*{Acknowledgements}

The authors would like to thank Stefano Camera for useful comments and questions during the development of this project. We would also like to thank Sourabh Paul for his comments on the final manuscript. \edittt{Lastly, we would like to extend our gratitude to the referee whose remarks improved the quality of the final paper.}

SC is supported by a UK Research and Innovation Future Leaders Fellowship grant [MR/V026437/1] and also acknowledges funding from the UK Research and Innovation Future Leaders Fellowship grant [MR/S016066/1]. MGS, YL and JW acknowledge support from the South African Radio Astronomy Observatory and National Research Foundation (Grant No. 84156). 
IPC acknowledges support from the `Departments of Excellence 2018-2022' Grant (L.\ 232/2016) awarded by the Italian Ministry of University and Research (\textsc{mur}).
AP is a UK Research and Innovation Future Leaders Fellow [grant MR/S016066/1].
LW is a UK Research and Innovation Future Leaders Fellow [grant MR/V026437/1].
MS acknowledges support from the AstroSignals Synergia grant CRSII5\_193826 from the Swiss National Science Foundation. 
PS is supported by the Science and Technology Facilities Council [grant number ST/P006760/1] through the DISCnet Centre for Doctoral Training.
This result is part of a project that has received funding from the European Research Council (ERC) under the European Union's Horizon 2020 research and innovation programme (Grant agreement No. 948764; PB). PB acknowledges support from STFC Grant ST/T000341/1. JF acknowledges support from the Fundação para a Ciência e a Tecnologia (FCT) through the Investigador FCT Contract No. 2020.02633.CEECIND/CP1631/CT0002 and the research grants UIDB/04434/2020 and UIDP/04434/2020.

We acknowledge the use of the Ilifu cloud computing facility, through the Inter-University Institute for Data Intensive Astronomy (IDIA). The MeerKAT telescope is operated by the South African Radio Astronomy Observatory, which is a facility of the National Research Foundation, an agency of the Department of Science and Innovation. 

For the purpose of open access, the author has applied a Creative Commons Attribution (CC BY) licence to any Author Accepted Manuscript version arising.

\section*{Data Availability}

\edittt{The data underlying this article will be shared on reasonable request to the corresponding author. Access to the raw data used in the analysis is public (for access information please contact archive@ska.ac.za).}



\bibliographystyle{mnras}
\bibliography{Bib} 



\bsp	
\label{lastpage}
\end{document}